\documentclass[final,5p,times,preprint]{elsarticle}

\usepackage{graphicx}
\usepackage{amssymb}

\journal{Journal of Magnetism and Magnetic Materials}

\begin{document}

\begin{frontmatter}

\title{Iron Based Superconductors: Pnictides versus Chalcogenides
}

\author[1,2]{M.V. Sadovskii\corref{cor1}}
\ead{sadovski@iep.uran.ru}
\author[1]{E.Z. Kuchinskii}
\author[1]{I.A. Nekrasov}

\address[1]{Institute for Electrophysics, Russian Academy of Sciences, Ural Branch,
Amundsen str. 106, Ekaterinburg, 620016, Russia}
\address[2]{Institute for Metal Physics, Russian Academy of Sciences, Ural Branch,
S. Kovalevskaya str. 18, Ekaterinburg, 620990, Russia}

\cortext[cor1]{Corresponding author.}

\begin{abstract}
We present a brief review of the present day situation with studies of 
high-temperature superconductivity in iron pnictides and chalcogenides. 
Recent discovery of superconductivity with T$_{c}>$ 30 K in  
A$_x$Fe$_{2-x/2}$Se$_2$ (A=K,Cs,Tl,…) represents the major new step in the 
development of new concepts in the physics of  Fe - based high-temperature 
superconductors.  We compare  LDA  and ARPES data on the band structure and 
Fermi surfaces of novel superconductors and those of the previously studied 
FeAs superconductors, especially isostructural 122 - superconductors like 
BaFe$_2$As$_2$. It appears that electronic structure of new superconductors 
is rather different from that of FeAs 122 - systems. In particular, no nesting 
properties of electron and hole - like Fermi surfaces is observed, casting 
doubts on most popular theoretical schemes of  Cooper pairing for these systems.  
Doping of novel materials is extremely important as a number of topological 
transitions of Fermi surface near the $\Gamma$ point in the Brillouin zone 
are observed for different doping levels. The discovery of  Fe vacancies 
ordering and antiferromagnetic (AFM) ordering at pretty high temperatures 
(T$_N>$ 500 K), much exceeding superconducting T$_c$ makes these  systems 
unique antiferromagnetic superconductors with highest T$_N$ observed up to now. 
This poses very difficult problems for theoretical understanding of 
superconductivity. We discuss the role of both vacancies and AFM ordering 
in transformations of band structure and Fermi surfaces, as well as their 
importance for superconductivity. In particular, we show that system remains
metallic with unfolded Fermi surfaces quite similar to that in paramagnetic state.
Superconducting transition temperature T$_c$ of new superconductors is 
discussed within the general picture of superconductivity in multiple band 
systems. It is demonstrated that both in  FeAs - superconductors and in new 
FeSe - systems the value of T$_c$ correlates with the value of the total density 
of states (DOS) at the Fermi level.

\end{abstract}


\end{frontmatter}

\section{Introduction}

Discovery of iron based high-temperature superconductors \cite{kamihara_08}
attracted a lot of scientific attention leading to a remarkable flow 
of experimental and theoretical works (for review see~\cite{UFN_90,Hoso_09}).
The main classes of iron (pnictides and chalcogenides) based 
superconductors known at the moment are:
\begin{enumerate}
\item{Doped RE1111 (RE=La,Ce,Pr,Nd,Sm,Tb,Dy) Fe pnictides with  
T$_c$ about 25--55 K, with chemical compositions like RE O$_{1-x}$F$_x$FeAs 
\cite{kamihara_08,chen,zhu,mand,chen_3790,chen_3603,ren_4234,ren_4283,TbDy}.} 
\item{Doped A122 (A=Ba,Sr), such as Ba$_{1-x}$K$_x$Fe$_2$As$_2$ 
\cite{rott,ChenLi,Chu,Bud} and T$_c$ about 38 K.}
\item{111 systems like Li$_{1-x}$FeAs with T$_c\sim$ 18 K~\cite{cryst,wang_4688}.}
\item{(Sr,Ca,Eu)FFeAs \cite{Tegel,Han} with T$_c\sim$ 36 K \cite{Zhu}.}
\item{Sr$_4$(Sc,V)$_2$O$_6$Fe$_2$(P,As)$_2$ with T$_c\sim$ 17 K \cite{42622}.}
\item{FeSe$_x$, FeSe$_{1-x}$Te$_x$ with T$_c$ up to 14 K \cite{FeSe}.}
\item{(K,Cs)$_x$Fe$_{2-y}$Se$_2$ and similar with T$_c$ up to 31K~\cite{Guo10,Krzton10}.}
\end{enumerate}
Among these, most recently discovered Fe chalcogenides like K$_x$Fe$_2$Se$_2$
and Cs$_x$Fe$_2$Se$_2$ with rather high values of superconducting
transition temperature: T$_c$=31K~\cite{Guo10} and 27K~\cite{Krzton10} were 
followed by $T_c=$31K in (Tl,K)Fe$_x$Se$_2$ \cite{Fang} and form apparently a distinct
new class.
These systems are isostructural to the FeAs 122 - systems \cite{UFN_90,Hoso_09}, 
while $T_c$ values for these Fe chalcogenides are a bit smaller than in similar pnictides. 
According to most recent data, in most of the chalcogenides 
Fe vacancies are intrinsic, so that most general chemical composition is usually 
written now as A$_x$Fe$_{2-y}$Se$_2$. In particular, Fe vacancy ordering was 
discovered in K$_{0.8}$Fe$_{1.6}$Se$_2$, and most strikingly this system seems to be
an ordered antiferromagnet (AFM) with pretty large Neel temperature 
(about 578K) \cite{Shermadini,Bao11}. Further ARPES investigations of these systems produced
experimental Fermi surface maps \cite{Mou11,Wang11,Zhao11} significantly 
different from previously studied for Fe pnictides~\cite{UFN_90,Hoso_09}.
Review of recent findings on Fe chalcogenide superconductors can be found
in Ref. \cite{Iva_Rev}.

In this work we discuss electronic structure, densities of 
states (DOS) and Fermi surfaces of Fe chalcogenides, as compared to Fe pnictides.
We present some estimates on the values of T$_c$, demonstrating that in all
Fe - based superconductors there is a definite correlation between T$_c$ and
the values of DOS at the Fermi level. Also we analyze the role of
Fe vacancies and AFM ordering in K$_x$Fe$_{2-x/2}$Se$_2$ in formation of its
electronic spectrum and Fermi surfaces, which are compared with available ARPES 
data.

\section{Electronic structure}

There are plenty of papers on LDA band structure of
La111 \cite{singh,dolg,mazin}, LaOFeP \cite{lebegue},
RE111 series \cite{Nekr}, BaFe$_2$As$_2$ \cite{Nekr2,Shein, Krell},
LiFeAs \cite{Nekr3,Shein2}, (Sr,Ca)FFeAs \cite{Nekr4,Shein3}, Sr42622 
\cite{Shein4}. Electronic structure of Fe(S,Se,Te) materials was discussed
in Ref.~\cite{SinghFeSe}. LDA calculations of electronic spectrum of newly
discovered (K,Cs)$_x$Fe$_2$Se$_2$ were described 
recently in~\cite{Shein_kfese,Nekr_kfese}.

To illustrate the general picture of the energy spectrum of Fe pnictides and chalcogenides
in the upper part of Fig.~1 we show LDA calculated total, Fe-3$d$ and As-4$p$ DOS (left panel) 
matched with the band dispersions (right panel) for typical representatives of 
both classes. From Fig.~1 one can see that around the Fermi 
level (from -2.5 eV to +2.5 eV) there are practically only Fe-3$d$ states present, while As-4$p$ 
states are contributing at lower energies (from -2.5 eV down to -6.0 eV).

\begin{figure}[h]
\includegraphics[clip=true,width=0.5\textwidth]{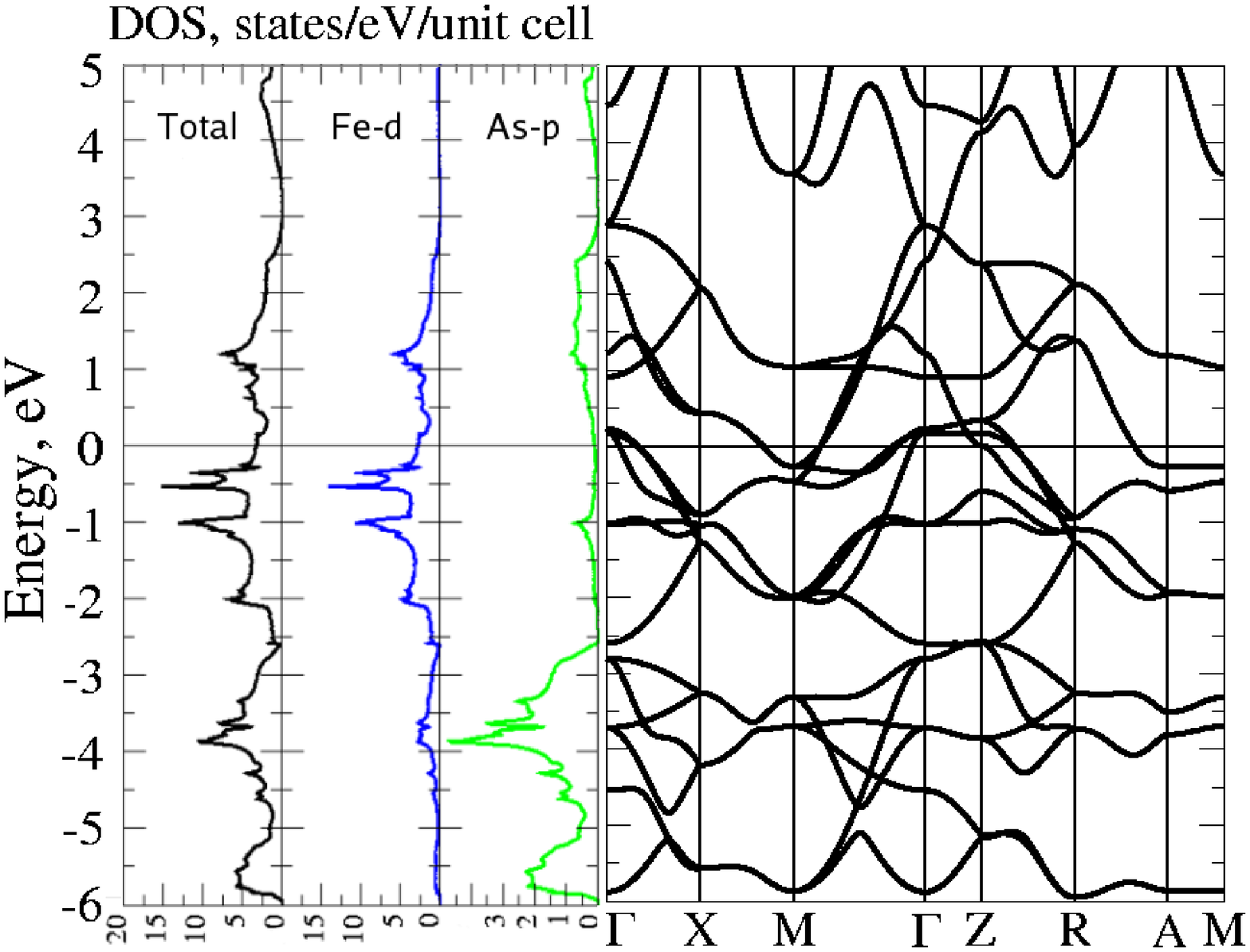}
\includegraphics[clip=true,width=0.5\textwidth]{K_DOS_bands.eps}
\label{fig1}
\caption{LDA calculated band dispersions and densities of states 
for typical representatives of iron based superconductors: 
LiFeAs (upper panel) for pnictides and KFe$_2$Se$_2$  (lower panel)
for chalcogenides. The Fermi level $E_F$ is at zero energy.} 
\end{figure}

In a bird eye (large-energy scale) view  K$_x$Fe$_2$Se$_2$ has similar
band dispersions as those in pnictides. However, there are some quantitative 
differences, e.g. all Fe-3d and Se-4p states in new systems are separated in energy
in contrast to Fe-3d and As-4p Ba122. Also Se-4p states are of about 0.7 eV 
lower than As-4p states. At the same time, similarly to pnictides the Fermi level $E_F$ 
in Fe chalcogenides is crossed only by Fe-3d states.

However, at lower energy-scale there is a major difference in spectra of 
both classes.
In Fig.~2 we compare LDA calculated electron spectrum in the immediate vicinity 
(relevant for superconductivity) of the Fermi level for both Ba122 \cite{Nekr2}
system and K$_x$Fe$_2$Se$_2$ \cite{Nekr_kfese}.
To some extent Ba122 bands near $E_F$ (upper part of Fig.~2) would match
those for KFe$_2$Se$_2$ if we shift them down in energy by 
about 0.2 eV. Main difference between old and new systems is seen around 
$\Gamma$ point. For KFe$_2$Se$_2$ systems antibonding part Se-4p$_z$ band in the 
Z-$\Gamma$ direction forms electron-like pocket. In Ba122 corresponding band 
lies about 0.4eV higher and goes much steeper, thus it is quite far away from 
$\Gamma$ point. However, if we dope KFe$_2$Se$_2$ systems 
(in a rigid band manner) with holes 
we obtain bands around $\Gamma$ point (close to the Fermi level) very similar 
to those in case of Ba122. Namely at 60\% hole doping we obtain Fermi surfaces
with three hole-like cylinders, while stoichiometric KFe$_2$Se$_2$ has one small 
electron pocket and larger hole like one near 
$\Gamma$ point. Thus, in fact under hole doping we expect several topological 
transitions of the Fermi surfaces \cite{Nekr_kfese}, which demonstrates potentially
rich effects of doping this system.

In general, the electronic structure of new chalcogenide superconductors close to
the Fermi level is significantly different from those of FeAs 122 systems. 
In particular no nesting of electron and hole - like Fermi surfaces is observed 
\cite{Nekr_kfese}, casting doubts on some of the most popular theoretical 
schemes of Cooper pairing developed for iron pnictides.

\begin{figure}[h]
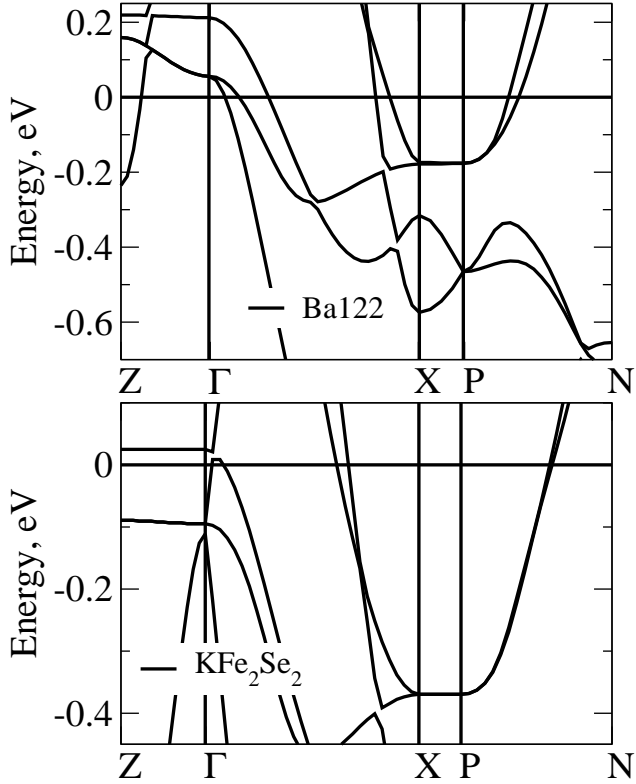

\includegraphics[clip=true,width=0.45\textwidth]{Ba122_Ef_bands.eps}
\includegraphics[clip=true,width=0.45\textwidth]{K_Ef_bands.eps}
\caption{Top panel -- LDA calculated band dispersions in the vicinity of the Fermi 
level for Ba122; Bottom panel -- KFe$_2$Se$_2$.
The Fermi level is at zero energy.} 
\end{figure}

\section{Anion height and DOS control T$_c$ ?}

It was discovered in Ref. \cite{hPn2} that superconducting temperature T$_c$
nonmonotonically depends on anion height $\Delta z_a$ with 
respect to Fe layer (see Fig.~3, triangles). Clear maximum is seen at 
about $\Delta z_a\sim$1.37\AA\  (see also Table~1).
Following this idea we performed systematic LDA computations
of total density of states N$(E_F)$ for number of iron based
superconductors (see Fig.~3, circles and Tab. 1) which have different 
$\Delta z_a$ \cite{Kucinskii10}. Here we also add similar results for
the new iron chalcogenides (K,Cs)Fe$_2$Se$_2$.
Nonmonotonous behavior of DOS can be explained by hybridization effects.
Namely, as a governing structural parameter characterizing hybridization 
strength one can chose $a$-Fe-$a$ angle --  an angle between anions ($a$) and 
Fe within the same tetrahedron. The value of this angle corresponding to the  
strongest hybridization is 109.45$^\circ$, i.e.  for an ideal anion tetrahedron 
with Fe in the very center of it.
Other crystal structure parameters are not changed very much from system to 
system and do not have any transparent dependence of $\Delta z_a$.

\begin{table}[htb]
\center
\footnotesize
\caption{LDA total DOS N(E$_F$), calculated and experimental T$_c$ 
for iron based superconductors.}
\label{tab1}
\begin{tabular}{|l|c|c|c|c|}
\hline
System                        &$\Delta z_a$, \AA & N(E$_F$),      & T$^{BCS}_c$, K &  T$_c^{exp}$, K\\
                              &                  & states/cell/eV &                &                \\
\hline
LaOFeP                        & 1.130            & 2.28           &    3.2         &     6.6        \\
Sr$_4$Sc$_2$O$_6$Fe$_2$P$_2$  & 1.200            & 3.24           &    19          &     17         \\
LaOFeAs                       & 1.320            & 4.13           &    36          &     28         \\
SmOFeAs                       & 1.354            & 4.96           &    54          &     54         \\
CeOFeAs                       & 1.351            & 4.66           &    48          &     41         \\
NdOFeAs                       & 1.367            & 4.78           &    50          &     53         \\
TbOFeAs                       & 1.373            & 4.85           &    52          &     54         \\
SrFFeAs                       & 1.370            & 4.26           &    38          &     36         \\
BaFe$_2$As$_2$                & 1.371            & 4.22           &    38          &     38         \\
CaFFeAs                       & 1.420            & 4.04           &    34          &     36         \\
CsFe$_2$Se$_2$                & 1.435            & 3.6            &    29          &     27         \\
KFe$_2$Se$_2$                 & 1.45             & 3.94           &    34          &     31         \\
LiFeAs                        & 1.505            & 3.86           &    31          &     18         \\
FeSe                          & 1.650            & 2.02           &     3          &     14         \\

\hline
\end{tabular}                                                                                                                                                           
\end{table}
\normalsize

To estimate superconducting critical temperature $T_c$ as a function of 
the $\Delta z_a$ one can apply the elementary BCS theory, where
in the expression $T_c=1.14\omega_D e^{-1/\lambda}$ corresponding
N$(E_F)$ enters into the dimensionless pairing interaction constant 
$\lambda=gN(E_F)/2$ ($g$ is the appropriate dimensional coupling constant).
Taking the Debye frequency $\omega_D$=350 K in rough accord with 
neutron scattering experiments on phonon density of states for La111 
\cite{Christianson} and Ba122 \cite{Mittal} systems, we can
find $g$ to fit the experimental value of $T_c$ for Ba122 system since this 
system possesses probably most stable value of $T_c$ (about 38 K) with respect 
to the way of sample preparation and doping. Thus we obtain the value of 
dimensionless coupling constant $\lambda$=0.43. Then just fixing the value of 
$g$ as for Ba122 we obtain $T_c$ values for all other systems, taking into 
account only the appropriate change of the density of states (Fig.~3, 
diamonds and Table 1). The agreement of these elementary estimates with
experimental values of T$_c$ is rather striking.

\begin{figure}[ht]
\includegraphics[clip=true,angle=270,width=0.55\textwidth]{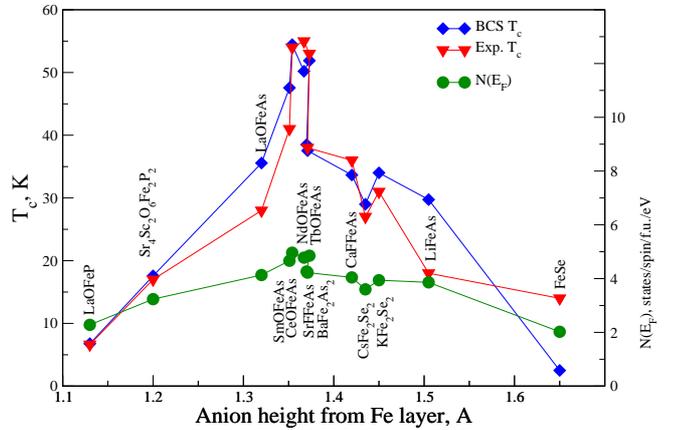}
\label{fig2}
\caption{LDA calculated total DOS values $N(E_F)$ (circles, right scale) 
and superconducting transition temperatures $T_c$ (left scale) obtained from 
simple BCS (stars) and  
experimental $T_c$ values (triangles) versus anion height $\Delta z_a$ over Fe 
layer for a number of iron based high temperature superconductors.} 
\end{figure}

In fact we do not adhere at the moment to any specific pairing mechanism, as
$\omega_D$ in BCS expression does not necessarily correspond to phonon
frequency, it can denote the characteristic frequency of any kind of Bosons
responsible for the pairing ``glue''. 
Our results only show unambiguous correlation between the values of
superconducting $T_c$ and those of the total density of electronic states at
the Fermi level for the whole class of iron based superconductors
(including the new chalcogenides), thus supporting the usual BCS-like pairing 
mechanism in these systems. The fit using more elaborate expression for T$_c$,
like e.g. Allen-Dynes formula, also produces rather satisfactory results
\cite{Kucinskii10}. Relatively high values of effective pairing couplings,
necessary to obtain experimentally observed values of T$_c$, can be understood as
due to multiple band electronic structure of new superconductors and importance
of inter - band couplings \cite{m_bands}. Special properties of electronic 
spectrum, such as widely popular ``nesting'' of electron and hole Fermi surfaces
are not necessary at all. At the same time, inter - band {\em repulsion} of any
kind, leading to $s^{\pm}$ - pairing seems preferable (non-phonon) mechanism of 
pairing, facilitating higher values of T$_c$ in both pnictides and chalcogenides.

\section{Vacancies and antiferromagnetism in KFe$_2$Se$_2$}

The recent discovery ~\cite{Shermadini,Bao11} of Fe vacancies ordering and
antiferromagnetic ordering at pretty high temperature
T$_N \sim$ 580 K, much exceeding superconducting T$_c$,
in the A$_x$Fe$_{2-x/2}$Se$_2$ (A=K,Cs,Tl,...) makes these
systems unique antiferromgnetic superconductors with
highest T$_N$ observed up to now. This poses some difficult problems
for theoretical understanding of superconductivity.
Here we discuss the role of both vacancies and AFM ordering
in transformation of band structure and Fermi surfaces.

Ordering of Fe vacancies in the K$_{0.8}$Fe$_{1.6}$Se$_2$ system \cite{Bao11}
provides $\sqrt{5}\times \sqrt{5}$ supercell. Corresponding ordered vacancies together
with antiferromgnetic order within the Fe-layer~\cite{Bao11} are presented in
Fig.~4. Translation vectors of the supercell because of AFM order are ${\bf a}_1$
and ${\bf a}_2$. Translation vectors corresponding to supercell with ordered vacancies
are  ${\bf b}_1$ and ${\bf b}_2$.

\begin{figure}
\includegraphics[clip=true,width=0.4\textwidth]{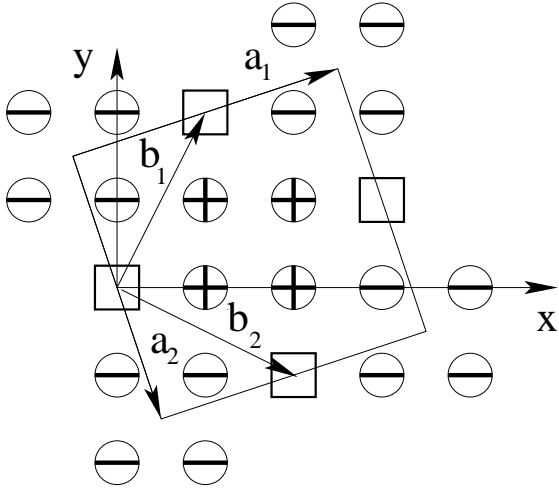}
\label{fig3}
\caption{Schematic picture of K$_{).8}$Fe$_{1.6}$Se$_2$ Fe-layer with vacancies (rectangle)
and experimentally observed AFM order (spin up -- circles with ``+'' inside, spin down --
circles with ``-'' inside). Translation vectors corresponding to AFM order are ${\bf a}_1$
and ${\bf a}_2$. Translation vectors for supercell with ordered vacancies are ${\bf b}_1$ and ${\bf b}_2$.} 
\end{figure}

We performed LSDA calculations for a simplified crystal structure of
K$_{0.8}$Fe$_{1.6}$Se$_2$ given in Ref.~\cite{Bao11}. Our results are similar but not
identical to other LSDA calculations on this system \cite{Cao11,Yan11}. For
example we obtain metallic AFM state with finite density of states at the Fermi
level, while energy gap forms at lower energies. Because of folding effects 
automatically included in LSDA code, it is rather difficult to make
direct comparison of LSDA band structure for K$_{0.8}$Fe$_{1.6}$Se$_2$ and its Fermi
surfaces with unfolded LDA bands and Fermi surfaces like those shown above.
Also Fermi surface maps from ARPES experiments \cite{Mou11,Wang11,Zhao11} 
are routinely shown in the unfolded Brillouin zone of
parent KFe$_{2}$Se$_2$, like that relevant for nonmagnetic LDA 
\cite{Shein_kfese,Nekr_kfese}. So these Fermi surfaces are quite difficult 
to compare with extremely folded LSDA ones \cite{Cao11,Yan11}.

To overcome this problem we use simplified  (semi)analytic model approach first 
used by us in Ref. \cite{Kuch_SP}. To this end we fit LDA bands of Ref.~\cite{Nekr_kfese} 
with simple parabolic bare spectra in the vicinity of the Fermi level using
the approach of Ref. \cite{Kuch_SP} 
and analyze multiple electron scattering (within iron plane) by vacancies (CDW) with potential
$V_1({\bf r})=2\Delta_1 (\cos {\bf  Q}_1 {\bf r} + \cos {\bf Q}_2 {\bf r})$
and AFM (SDW) potential $V_2({\bf r}) = 2\Delta_2 (\cos {\bf X}_1 {\bf r} + 
\cos {\bf X}_2 {\bf r})$.
Corresponding vacancy CDW and SDW vectors are ${\bf Q}_1 = 2\pi(0.4,0.2)$, 
${\bf Q}_2 = 2\pi(-0.2,0.4)$, ${\bf X}_1 = 2\pi(0.1,0.3)$, ${\bf X}_2 = 2\pi(-0.3,0.1)$.

Initial (retarded) bare Green function is taken as:
\begin{equation}
g^{ij}({\bf k})=g^i\delta_{ij}=\frac{1}{E-\epsilon_i ({\bf k})+i\delta}\delta_{ij}
\label{G0}
\end{equation}
where $i,j$ are band indices, $\epsilon_i ({\bf k})$ --  $i$-th band model
electronic spectrum.
Because of multiple scattering on vacancies (CDW) and AFM (SDW) order, band electron 
with momentum ${\bf k}$ could be scattered by any of ten possible momenta 
(See Table 2 of scattering vectors summation).
Namely, electron can
preserve its momentum, or change it in nine other ways, being scattered by
${\bf Q}_l$; $\bar {\bf Q}_l=-{\bf Q}_l$ ($l=1,2$); 
${\bf X}_l$; $\bar {\bf X}_l=-{\bf X}_l$ ($l=1,2$); ${\bf Y}=(\pi,\pi)$.
Thus, to find one-band diagonal Green's function
$G({\bf k},{\bf k})\equiv G$ and 
nine off-diagonal: 
$G({\bf k}\pm {\bf Q}_l,{\bf k})\equiv F_{l(\bar l)}$, $G({\bf k}\pm {\bf X}_l,{\bf k})\equiv \Phi_{l(\bar l)}$
 and $G({\bf k}\pm {\bf Y},{\bf k})\equiv \Psi$, 
we end up with system of 10 linear equations.
Such approach can also be generalized to the multiple band case in a simplified
way similar to that used in Ref.~\cite{Kuch_SP}, assuming that for both vacancy 
and SDW scattering both inter- and intraband scattering amplitudes are identical. 
Thus we can obtain for the multiple case the system of ten linear equations for
following Green functions -- $G^{ij}$, $F_{l(\bar l)}^{ij}$, $\Phi_{l(\bar l)}^{ij}$, $\Psi^{ij}$,
which now have two band indices. The equation for diagonal Green function is written as:
\begin{equation}
G^{ij}=g^i\delta_{ij}+g^i\sum_{l=12\bar 1\bar 2}[\Delta_1\sum_m F_l^{mj}+\Delta_2\sum_m \Phi_l^{mj}].
\label{1eq_sys}
\end{equation}
Once we sum up over $i$ we come to
\begin{equation}
G^j=g^j+g\sum_{l=12\bar 1\bar 2}[\Delta_1F_l^j+\Delta_2\Phi_l^j].
\label{sum1eq}
\end{equation}
The rest of other nine equations for $G^j=\sum_iG^{ij}$, $F_l^j=\sum_iF_l^{ij}$, 
$\Phi_l^j=\sum_i\Phi_l^{ij}$, $\Psi^j=\sum_i\Psi^{ij}$
can be obtained using Table 2 of scattering vectors summation.
More details on this approach will be presented in Ref. \cite{geks}, devoted to
similar model of hexagonal transition metal dichalcogenides, like NbSe$_2$ and
TaSe$_2$.

Finally by solving this system of linear equations we can get the
diagonal Green's function $G^{ij}({\bf k},{\bf k})$ ($i,j$ -- band indices) and
define spectral functions
\begin{equation}
A(E,{\bf k})=-\frac{1}{\pi}Im\sum_{i}G^{ii}({\bf k},{\bf k}).
\label{SpDen}
\end{equation}
To account for finite experimental spectral resolution we broaden all 
our results by substituting $E\to E+i\gamma$, with $\gamma = 0.03 eV$, 
corresponding to typical ARPES resolution.

\begin{table}
\label{Tab2}
\caption {Table of scattering vectors summation.}  
\begin{tabular}{|c||c|c|c|c|}
\hline
&  ${\bf Q}_1$ & ${\bf Q}_2$  & $\bar {\bf Q}_1$ & $\bar {\bf Q}_2$ \\  
\hline \hline
${\bf Q}_1$ &  ${\bf Q}_2$ & $\bar {\bf Q}_2$ & 0 & $\bar {\bf Q}_1$ \\ 
\hline 
${\bf Q}_2$ & $\bar {\bf Q}_2$  & $\bar {\bf Q}_1$ & ${\bf Q}_1$ & 0 \\ 
\hline
$\bar {\bf Q}_1$ & 0 &  ${\bf Q}_1$ & $\bar  {\bf Q}_2$ & ${\bf Q}_2$ \\
\hline
$\bar {\bf Q}_2$ & $\bar {\bf Q}_1$ & 0 &  ${\bf Q}_2$ & ${\bf Q}_1$ \\
\hline
\end{tabular}
\hspace{0.5cm}
\begin{tabular}{|c||c|c|c|c|}
\hline
&  ${\bf X}_1$ & ${\bf X}_2$  & $\bar {\bf X}_1$ & $\bar {\bf X}_2$ \\  
\hline \hline
${\bf X}_1$ &  $\bar {\bf Q}_2$ & ${\bf Q}_2$ & 0 & ${\bf Q}_1$ \\ 
\hline 
${\bf X}_2$ & ${\bf Q}_2$  & ${\bf Q}_1$ & $\bar {\bf Q}_1$ & 0 \\ 
\hline
$\bar {\bf X}_1$ & 0 &  $\bar {\bf Q}_1$ & ${\bf Q}_2$ & $\bar {\bf Q}_2$ \\
\hline
$\bar {\bf X}_2$ & ${\bf Q}_1$ & 0 &  $\bar {\bf Q}_2$ & $\bar {\bf Q}_1$ \\
\hline
\end{tabular}
\hspace{0.5cm}
\begin{tabular}{|c||c|c|c|c|}
\hline
&  ${\bf Q}_1$ & ${\bf Q}_2$  & $\bar {\bf Q}_1$ & $\bar {\bf Q}_2$ \\  
\hline \hline
${\bf X}_1$ & ${\bf Y}$ & $\bar {\bf X}_1$  & ${\bf X}_2$ & $\bar {\bf X}_2$ \\ 
\hline 
${\bf X}_2$ & ${\bf X}_1$ & ${\bf Y}$ & $\bar {\bf X}_2$ & $\bar {\bf X}_1$ \\ 
\hline
$\bar {\bf X}_1$ & $\bar {\bf X}_2$ & ${\bf X}_2$ & ${\bf Y}$ & ${\bf X}_1$ \\ 
\hline
$\bar {\bf X}_2$ & ${\bf X}_2$ & ${\bf X}_1$  & $\bar {\bf X}_1$ & ${\bf Y}$ \\
\hline
\end{tabular}

\vspace{.5cm}
\begin{tabular}{|c||c|c|c|c|c|c|c|c|}
\hline
& ${\bf Q}_1$ & ${\bf Q}_2$ & $\bar {\bf Q}_1$ & $\bar {\bf Q}_2$ & ${\bf X}_1$ & ${\bf X}_2$ & $\bar {\bf X}_1$ & $\bar {\bf X}_2$ \\  
\hline \hline
${\bf Y}=(\pi ,\pi )$ & $\bar {\bf X}_1$ & $\bar {\bf X}_2$ & ${\bf X}_1$ & ${\bf X}_2$ & $\bar {\bf Q}_1$ & $\bar {\bf Q}_2$ & ${\bf Q}_1$ & ${\bf Q}_2$ \\ 
\hline 
\end{tabular}
\end{table}

The values of model scattering amplitudes $\Delta_1$ and $\Delta_2$ and
chemical potential were obtained by approximate fitting DOS obtained from our 
model calculations (solid line in Fig.~5) to (similarly broadened) LSDA DOS 
(dashed line in Fig.~5) in a narrow energy interval close to the Fermi level. 
It can be seen that this fit is rather good in this energy interval, where 
parabolic (two-dimensional) fit for energy bands is also satisfactory.

\begin{figure}
\includegraphics[clip=true,width=0.5\textwidth]{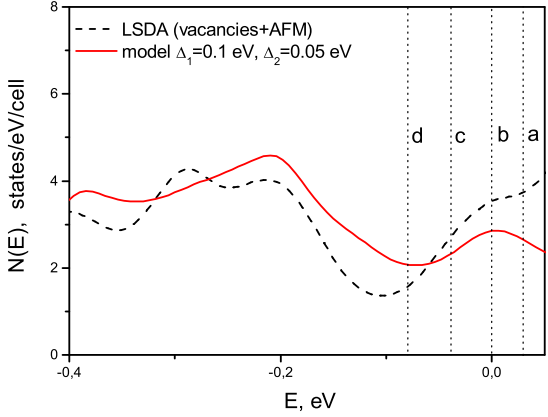}
\label{fig4}
\caption{Comparison of part of total LSDA DOS (dashed line) for 
K$_x$Fe$_{2-x/2}$Se$_2$ with model DOS (solid line) obtained from parabolic 
bare bands with vacancy (CDW) and AFM (SDW) scattering taken into account. 
Fermi level is zero. Letters a,b,c and d denote different doping levels,
corresponding to different Fermi surfaces shown in Fig.~7.} 
\end{figure}

In Fig.~6 model spectral function map (\ref{SpDen}) for K$_x$Fe$_{2-x/2}$Se$_2$ is 
presented. As mentioned above the bare spectra in the vicinity of the Fermi 
level were modelled by a number of parabolas (cf. Ref. \cite{Kuch_SP}). 
The general broadening (finite life time effects) comes out from finite $\gamma$
resolution effects. Quasiparticle bands with additional shadow like features 
due to vacancy and AFM ordering are clearly seen and can be observed in ARPES
experiments. These bands are to be compared with simple LDA results of 
Ref.~\cite{Nekr_kfese}, shown in the lower part of Fig. 2.

\begin{figure}
\includegraphics[clip=true,width=0.5\textwidth]{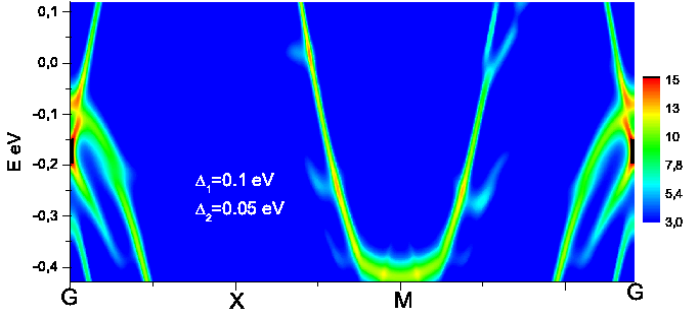}
\label{fig5}
\caption{Model spectral function map for for K$_x$Fe$_{2-x/2}$Se$_2$
obtained from parabolic bare bands with
vacancy (CDW) and AFM (SDW ) scattering taken into account. 
Fermi level is zero.} 
\end{figure}

\begin{figure}
\includegraphics[clip=true,width=0.5\textwidth]{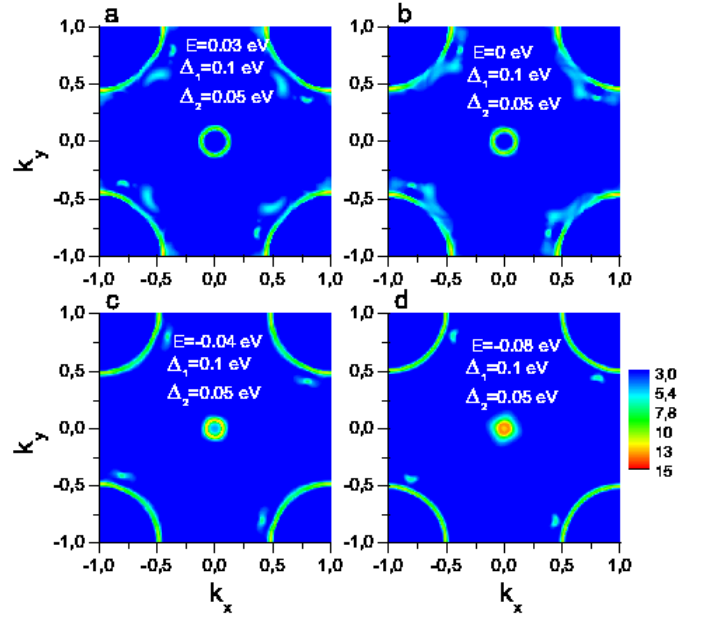}
\label{fig6}
\caption{Model Fermi surface maps for K$_x$Fe$_{2-x/2}$Se$_2$
obtained from parabolic bare bands with
vacancy (CDW) and AFM (SDW) scattering taken into account. Panels a,b,c and d
correspond to doping levels shown on Fig.~5.}
\end{figure}

Model Fermi surface maps for K$_x$Fe$_{2-x/2}$Se$_2$ with
vacancies and AFM (SDW) taken into account for different
doping levels are shown in Fig.~7. Panel (a) of Fig.~7
corresponds to electron doped case, panel (b) -- undoped and
panels (c) and (d) -- hole doped case. Corresponding
doping levels are displayed also on Fig.~5.
These (model) theoretical results are very similar to
experimental data  of Refs.~\cite{Mou11,Wang11,Zhao11}, with
rather large (electronic) cylinders around ($\pi,\pi$)-point
and rather small or even almost completely smeared out cylinders around 
$\Gamma$-point were observed. Similar picture was found also for parent 
KFe$_{2}$Se$_2$ system in simple LDA \cite{Nekr_kfese}.
Presence of vacancy (CDW) and AFM (SDW) scattering leads only to formation of
some low intensive shadow Fermi surfaces which would be rather hard to detect 
in ARPES experiments. In fact, most ARPES investigations of these system
demonstrate experimental Fermi surface maps quite similar to those shown in
Fig. 7 \cite{Mou11,Wang11,Zhao11}. 

The most important conclusion is that, rather unexpectedly, the role of vacancy
scattering and antiferromagnetic ordering in the formation of electronic spectrum in
the vicinity of the Fermi level is rather minor. The system remains metallic with 
well defined Fermi surfaces, allowing the formation of superconducting state on 
the background of AFM and vacancy order \cite{Bul_Buz}.

\section{Conclusion}

In this work we present comparative study of iron based
pnictide and chalcogenide superconductors. It was shown
that for both families superconducting transition temperature is apparently
controlled by anion height $\Delta z_a$  with respect to Fe plane, which is
directly correlated with the values of DOS at the Fermi level.
Unfolded Fermi surfaces observed in the ARPES experiments even in strong 
antiferromagnet K$_{0.8}$Fe$_{1.6}$Se$_2$ with ordered Fe vacancies are very similar to
those of parent (nonmagnetic) KFe$_{2}$Se$_2$. The system remains metallic
despite AFM and vacancy ordering and superconducting state with rather high 
T$_c$ forms on this unusual background.

\section {Acknowledgements} 

This work is partly supported by RFBR grant 11-02-00147 and was performed
within the framework of programs of fundamental research of the Russian 
Academy of Sciences (RAS) ``Quantum physics of condensed matter'' 
(09-$\Pi$-2-1009) and of the Physics Division of RAS  ``Strongly correlated 
electrons in solid states'' (09-T-2-1011).

\end {document}